# Autonomous Wireless Systems with Artificial Intelligence

Haris Gacanin, Nokia Bell Labs

*This paper discusses technology and opportunities to embrace artificial intelligence (AI) in the design of autonomous wireless systems. We aim to provide readers with motivation and general AI methodology of autonomous agents in the context of self-organization in real time by unifying knowledge management with sensing, reasoning and active learning. We highlight differences between training-based methods for matching problems and training-free methods for environment-specific problems. Finally, we conceptually introduce the functions of autonomous agent with knowledge management.*

## Introduction

The fifth generation (5G) wireless infrastructure aims to support heterogeneous service designs with diverse latency and reliability requirements as illustrated in Fig. 1. Consider these service design examples with interesting challenges: 1) massive Internet-of-Things, where devices, sensors and actuators give rise to the problem of network planning in real time; 2) broadband wireless leads to problems with real time radio resource management; 3) ultra-reliable communications require support of real time (i.e. simultaneous) adjustments for wireless infrastructure to latency and reliability in the orders of 99,99999%. With such designs of wireless systems, complexity and environment dynamics rises, radio resources are scattered, and diversity of system elements increases [1]. In 5G systems, simultaneous transition between such heterogeneous service types is not feasible. For example, with dedicated design the one can support low-latency communications (e.g. 1 ms latency) accompanied by low-rate control messaging, but low-latencywith high-rate applications is not feasible. The next-generation wireless systems are expected to autonomously utilize technologies in distributed fashion to satisfy simultaneous delivery of services with stringent requirements which arise in a dynamic way as illustrated in Fig. 1.

Design of autonomous wireless systems with simultaneous service delivery in real time cannot be accomplished by incremental changes to the present deterministic control and optimization methodologies. It requires a fundamental leap in the system's thinking by embedding machine intelligence into temporal wireless infrastructure itself. This means that the infrastructure will become aware of the way it is being used to anticipate actual requirements at the specific moment and what it is likely to be required at later time. As such it will facilitate wireless as a true application-aware platform for a plethora of novel applications.

We are now several years into explosion of machine learning (ML) in wireless networks, used to enrich decision-making by finding structures in data with ML – *knowledge discovery* – as means to describe the system performance [2]. Within the field of artificial intelligence (AI), ML evolved from computational learning theory as an efficient way to enable machines to process and learn from given data with little or no guidance at all. The automation of traditional SON employs deterministic decision-making methods to control coordination of self-organizing network (SON) functions [3]. Current data-driven methods in wireless systems with ML improve decision-making for traditional SON with applications in signal detection and channel estimation for massive multiple-input multiple-output systems, user association and spectrum sensing in cognitive radio; all leading to high-dimensional search-problems [4]. Moreover, learning is considered for SON to address mobility load balancing, mobility robustness optimization, coverage and capacity optimization, inter-cell interference coordination, random access channel optimization, cell outage detection to name just a few [5].

By looking beyond recent data-driven paradigm with ML [6]-[10], this forward-looking paper elaborates the concept of machine intelligence in wireless systems. Machine intelligence employs broader disciplines of AI such as *sensing, reasoning, active learning, and knowledge management* [11], [12]. With AI intelligent systems can be designed to perform "autonomous" tasks (e.g. planning, problem solving) without explicitly programmed to accomplish a single (repetitive) task, while being adaptive to different environments. Such system requires full awareness of its environment in real time with design not only through data-driven methods by ML but employing knowledge management by AI. Unlike automation efforts with traditional SON, this paper moves forward and introduces the methodology for autonomous operation (i.e. self-organization). First, we summarize the properties of training-free and training-based methods in wireless environment, while focusing on reinforcement learning (RL) as a major representative of AI. The principles of knowledge management are highlighted and then, we present a functional example of autonomous agent by synthesizing sensing, reasoning and active learning with knowledge management.

## Data-driven Self-organizing Network

These are the traditional SON functions [3]: 1) self-configuration (e.g. learning of configuration parameters, neighbors), 2) self-optimization (e.g. learning in mobility, load, handover, interference, capacity and coverage optimization) and 3) self-healing (e.g. fault analysis, detection). The

relationship of the traditional SON and AI methods is described in Table 1 with a short summary as follows:

*Self-configuration*: System configuration setup either on initial deployment depending on the current critical situation in terms of system operations: cell coverage and deployment, neighbor cell list, etc. Currently using deterministic AI methods provided by operator's auto-configuration server or placed locally on memory as system configuration backup.

*Self-optimization*: Deterministic rule-based system with automated parameter optimization according to global objectives: quality of service, capacity/bandwidth, coverage, etc. AI methods are characterized by deterministic states determined by logical formulas having constrained optimizer functions. Decision-making is deterministic with rule-based logical functions not supporting learning.

*Self-healing*: The system checks and adapts configurations in real time. AI methods are characterized by deterministic or probabilistic reasoning. Optimizer employs closed-loop learning or heuristics/meta-heuristics due to problem complexity supported by learning. Decision-making employs learning depending on feedback as supervised, semi-supervised or unsupervised.

*Related Works*

Helpful discussions on applications of AI to network management and orchestration are presented in [1]. In this work, a conceptual example of high-level RL framework for traffic-aware energy management has been presented. The framework is described as an abstract layer interacting with network elements (i.e. radio access network, virtual nodes, etc.) using open-source interfaces. In principle, ML is employed to design a learning function to enhance predictions or decision-making [4]. In [5], the authors discuss up to date research concepts including different learning and decision-making techniques (Genetic algorithms, Swarm intelligence, Neural networks, Fuzzy systems, Markov and Bayesian games) for traditional SON with deterministic coordination functions. An inspiring work on network planning using the concept of knowledge discovery is discussed in [6]. The authors present how the different planning and operation procedures can exploit discovered knowledge by data mining over the collected data sets. The limiting factor is the knowledge exploitation that is accomplished by deterministic tools, e.g. modelling of traffic for prediction, while decision-making is not designed to manipulate with discovered knowledge.

The early works on data-driven optimization and design in communications with ML dates to [13], where the scope of the cognitive radio research is extended towards more holistic approach. The authors present a framework for cognitive resource manager enabling autonomic local optimization methods of the communication stack, instead of focusing solely on the spectrum problem. While still having deterministic relationships between cognitive managers they may be used to federate individual cognitive radios to offer a systematic approach as a SON with distributed cross-layer optimization.

A cognitive optimization framework is presented in [8] for the network virtualization problems. The framework aims to automate a system level operation by learning-architecture functions with deep learning and probabilistic generative models. An insightful approach to learn optimization policy for network management problems by deep RL is presented in [10].

A recent successful attempt to shift from the knowledge-discovery toward knowledge-driven (autonomous) operation has been made in [11], where an autonomous agent is presented to address the problem of self-deployment of non-stationary radio nodes. In [12], the authors present a wireless environment-specific RL agent with Q-learning to solve a self-optimization problem with joint channel association and location optimization by retaining and reusing past experiences to reason out new optimization strategy.

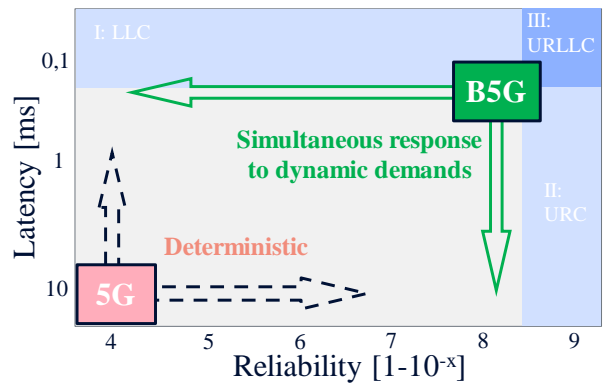

*Figure 1. Service delivery with heterogeneous requirements. Type I - LLC: Low-latency communications; Type II - URC: Ultra-reliable communications; Type III - URLLC: Ultra-reliable low-latency communications;*

## Training-based and Training-free AI Methods in Wireless Environment

A wireless environment within the context of AI has the following challenging properties:
- stochastic with infinite horizon;
- partially observable and dynamic;
- multiagent;
- episodic and continuous.

The applicability of AI in such environment is determined with a goal to predict or to operate autonomously. Under constraint of real time wireless content delivery, instead of following traditional classification on supervised, unsupervised and RL, we take intuitive approach and classify on training-based (prediction with supervision) and training-free (autonomy with reinforcement) methods in Fig. 2.

*Learning with Supervision*

Most of the recent works consider training-based learning for matching (i.e. prediction) problems. Examples, of such works are utilization of big data in radio, core and network management [14], [4]-[6]. Here, offline data with deterministic processing is employed to train prediction models by means of neural networks for example. For detailed taxonomy of big data utilization in SON for wireless networks please refer to [4].

Some noteworthy surveys on AI with supervised ML for decision-making in wireless systems discuss different techniques such as Genetic algorithms, Deep learning, Swarm intelligence, Neural networks, Fuzzy systems, Markov and

Bayesian games [4]-[7]. Clearly, for such approaches data is a key for learning. Recently considered deep learning requires a large amount of data to generalize that may pose difficulties in radio optimization with time-sensitive constraints (e.g. planning and optimizing a massive network of macro, micro and small cells within a limited spectrum band will be a challenge). Unlike computer vision, for time-sensitive applications such as Industrial automation or low-latency radio access in real time the wireless data is expected to be available in bursts and dynamic. Such scenario poses limitations on the model-based solutions with training since a model generalization is highly sensitive to limited experience (small training set). The impact and limitations of training for automation of wireless networks employing big data is relevant and ongoing research study [14].

On the contrary, this work advocates the need for solutions supporting autonomous wireless operations adaptive to changing environment in real time.

| Matching problems | Dynamic problems |
|---|---|
| **Training based** $p \quad a$ | **Training free** |
| Direct feedback | Reinforcement |
| Offline decision-making limited by training | Online decision-making limited with starting point |
| Cannot generalize from limited data | Cumulative reward exploitation |
| Sensitive to limited experience | Random exploration |
| Sensitive to data labeling and cleaning (by humans) | Knowledge domain needed |

*Figure 2. Training-based vs training-free properties for different types of problems; p = prediction and a = autonomy.*

|  | $a_1$ | $a_2$ | ... | $a_n$ |
|---|---|---|---|---|
| State 1 | - | 10 | 5 | 0.2 |
| State 2 | 100 | 7 | - | 1 |
| ... | ... | ... | ... | ... |
| State m | 2 | - | 30 | 5 |

*Table 2. Simple state-action-reward table where the numeric values represent estimated Q-values.*

### Learning by Acting

The other extreme deals with dynamic problems where the system operates in an unknown environment [9]. Training-free methods, represented by RL, autonomously build and exploit knowledge to learn a (near-)optimal policy of its own actions by using a reward as illustrated in Fig. 3. The agent can be designed as a passive learner where the policy is fixed, and the task is to learn the action values and learn a model of environment, or an active learner where the agent learns what to do by exploring to experience and learn how to behave.

Q-learning RL (Q-RL) is a model-free iterative technique with a goal to reach an optimal policy without needing to know their outcomes. Q-RL agent learns a table of Q(s,a) values (rewards or action values) being estimated reward of taking an action *a* in each state to keep information which actions have the highest reward. For example, consider a simple multi-node mesh wireless network as *k*-armed bandit problem of maximizing throughput, which has *m* states (defined by interference, collisions, load, etc.) and *n* possible actions *a* (defined as the system operating frequencies per node). Table 2 represents action-state space, where Q-values represent achievable user throughput. We denote the action selected on time step *t* as $a_t$, and the corresponding reward as $r_t$. The expected reward $q^*(s,a)=E[r_t \mid a_t=a, s_t]$ given that *a* is selected. By knowing Q-value per action, the solution is trivial as we always greedily select the action with highest value given by $a$ = arg max$_{a'}Q(s,a)$. Such selection always exploits what it knows to maximize immediate reward without trying new actions to check if they have higher throughput (Q-) values. Practically, $Q(s,a)$ are not known with certainty, but measured or estimated to be close to $q_t^*(a,s)$ with update by $Q(s,a) \leftarrow Q(s,a)+\alpha[r_t+\gamma$ max$_{a'}Q(s',a')-Q(s,a)]$, where α represents a learning rate while γ discounts the delayed reward impact.

Exploring the action space $\{a_1, a_2,... a_n\}$, the agent selects different frequency channels to improve Q-values of the non-greedy actions. The agent is greedy with probability 1-ε, while trying actions with random (Boltzmann or Gibbs) probability ε independently of the Q-values. While the number of steps increases to limit, each action is randomly selected an infinite number of times, ensuring that $Q(s,a)$ converges. This pose a challenge with wireless services in real time. In the wireless example above, such exploration requires radio nodes to switch to *n* frequency channels leading to user disconnections and service disruption [12].

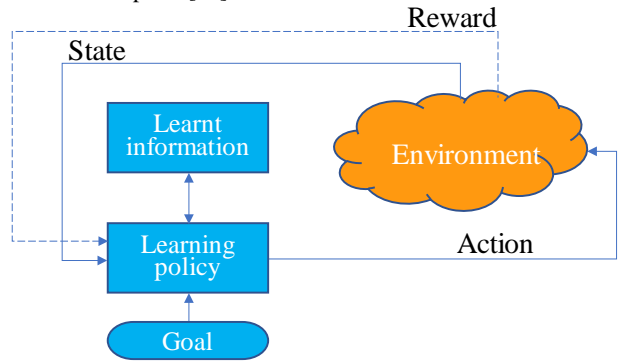

*Figure 3. Illustration of reinforcement agent interactions with environment and corresponding elements.*

Depending on the problem state-action space Table 2 representation of the Q-function may not be practical as the table size increases exponentially. However, for wireless environment-specific problems such representation may still be efficient with low memory requirements [12]. This is because radio access problems require fast response and may be efficiently characterized by a few data. Instead, neural networks may be used to approximate the Q-function, but this approach introduces training and requires mass of data that may pose limitations for wireless applications in real time.

*Sample RL applications*: Clearly, while the selection and/or combination of AI methods is a design choice, a strict wireless service and system requirements lead to different applications properties listed in Fig. 2. However, the challenge of balance between exploration and exploitation in RL is not clearly discussed. The exploration in wireless systems has negative effect on service delivery in real time, e.g. by exploring the radio node changes operating frequency that disrupts services. Helpful discussions in [2] presented applications of high-level RL framework for traffic-aware energy management using open-source interface, while an insightful approach to learn optimization policy for network management problems by deep RL is presented in [10].

*Open AI Tools*: Recent open tools such as TensorFlow, Scikit-learn, Microsoft Cognitive Toolkit, Torch, to name just a few, allow development of learning frameworks such as deep neural networks and other computational models. While these tools are intended for large-scale data manipulation on systems running optimally on both central processing units and graphics processing units, they should be re-designed for radio environment and real time applications with high reliability or low latency or having modest processing capabilities on user or radio nodes.

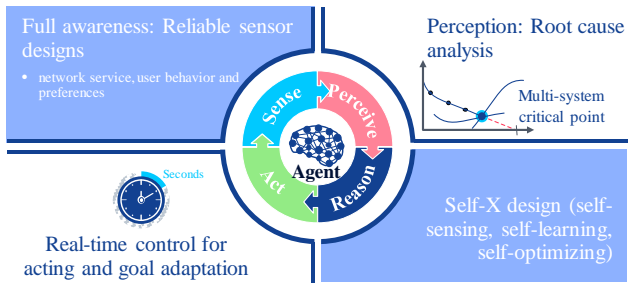

*Figure 4. Conceptual agent functions for autonomy with AI.*

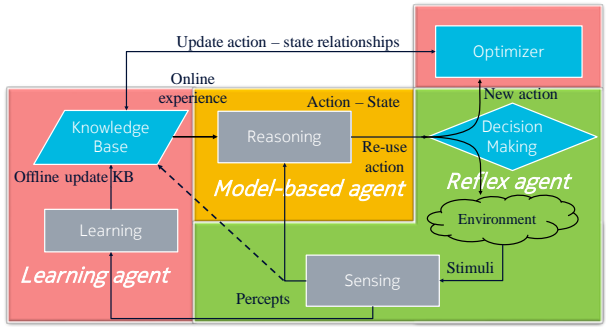

*Figure 5. Knowledge-driven intelligent agent design.*

**Knowledge-driven Operation with AI**

Self-organization is an autonomous process where a system's structure and functionality at the global level emerge from interactions among the "lower-level" subsystems without any external or centralized control. The subsystems interact either by means of direct communication or environment percepts without reference to the global goal. Figure 4 illustrates the autonomous agent needs to be fully aware of different network and service functions, while taking decisions and actions in real time based on predicted service sessions. From the agent's perspective, an established communication session is fully observable, but the system and user environments are partially observable. Contrary to data-driven (training-based) methodology, the new paradigm is needed to employ *knowledge-driven methodology* inspired by AI in real time.

The AI can be implemented in different forms such as rule-based system (RBS), ontology-based system (OBS), case-based reasoning (CBR), among others [15]. The RBS comprises a set of rules with predefined actions created by experts in the domain. Similarly, OBS applies logic-based reasoning for the domain attributes. Both RBS and OBS require explicit domain knowledge to define the relations between rules and actions or objects. Contrary, CBR relies on the system memory defined by knowledge base to build the knowledge using observations about previous actions and their impact on the system. As discussed in [12], an RL benefits from such reasoning system which significantly speeds up learning of unknown environment and improves the agent efficiency. The agent perceives its environment roughly through a sequence of sensing, reasoning and acting to build its own knowledge and use it in the future actions. Thus, good actions, e.g. that achieve target quality-of-service, can be reused directly in the future when similar network conditions are sensed, while bad actions, e.g. that create coverage holes, will initiate new action search.

We note here that traditional SON relies on deterministic coordination to handle network operations and management systems as pointed out in [3]. On the contrary, self-organization with AI aims to introduce autonomous decision-making by using knowledge management and active learning.

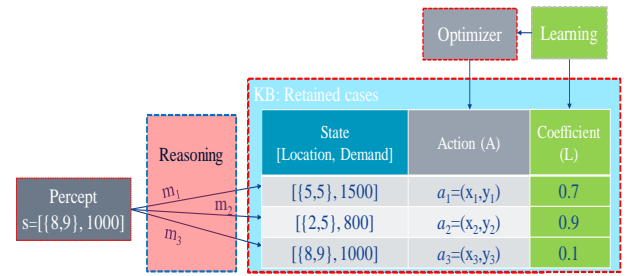

*Figure 6. A conceptual example of agent functions.*

*Agent Functions*

Figure 5 illustrates the intelligent agent design with knowledge management. The implementation design of the functions is flexible to support distributed wireless infrastructure. Such agent could be considered for single or multi active agent problems. Depending on design the implementation may be distributed, hybrid [12] or centralized between the nodes and cloud [10].

*Knowledge Base* (*KB*): KB represents of all the agent's knowledge acquired through interactions with an environment. The agent applies the following four stages on the KB:
- Retrieve the most relevant case to the current percept;
- Reuse the retrieved case to solve the perceived problem;
- Revise the KB by updating the actions or learning coefficients;
- Retain the new cases to be used in the future.

The agent functions illustrated in Fig. 5 implement the above four stages, while an implementation example is shown in Fig. 6. For instance, KB stores the knowledge as a triplet of a state, an action and a learning coefficient in the form represented in Fig 5. The percept vector refers to the current state provided by the Sensing function, while corresponding action with the learning coefficient defines the degree of how much the action impacts the performance as explain next. Together with Learning and Optimizer functions, the sensing provides inputs to the KB by percepts as indicated in the illustrative example in Fig. 6. More comprehensive design of the KB is described in [12].

*Sensing*: Sensing function in Fig. 5 is responsible for collection of network stimuli by means of measurements in real time via programmable interfaces. Some examples of measurements are modulation and coding schemes, load, channel state information, frequency, power level, channel busy time, re-transmissions, failed packets, sent packets, user application and location, to name just a few. Sensing function transforms stimuli into percepts and need to handle the following:
- Incomplete data for relevant periods of time;
- Spatially separated data may represent different structures;
- I. i. d data assumption does not hold;
- Extreme-large training needed for long-term supervised learning.

In the following contextual example[1], let us consider the problem of location optimization for non-stationary radio nodes (e.g. drones, robots). Depending on design choice percepts can be [11]: Node location, Location allowed or not; User association or not with an access point; Per user throughput and/or demand, etc. The percepts define the system-environment state to detect when the current configuration becomes sub-optimal and signals to evaluate the current operational state of the network. This is done for each radio node for two successive sensing samples to detect unsatisfactory state. In other example, a state is perceived by achievable throughput that is available at nodes and users. Some other examples are user battery status, mobility, coverage/signal quality indicator and achievable throughput depending on design problem. Note that proper data normalization and scaling of vectors is needed.

*Reasoning*: Reasoning manipulates with entries of the KB to search for an action by identifying similarity between cases in KB and newly observed percept [11]. CBR with KB implements AI, where the reasoning function searches for the best action in response to perceived state. In the example of location optimization, Fig. 6 illustrates the agent implementation with percept *s*=[Location, User demand] per node. When user throughput is lower than its demand, the function compares the current percept with previously stored states in KB to calculate a similarity factor *m* implemented as minimum distance Euclidean classifier, Bayesian classifier, etc. In the case of high similarity, e.g. $m_3$, while learning coefficient *L* is low indicating a poor location action in the past, the new action $a_3$ is generated by the optimizer and updated in KB by learning function as discussed below. The coefficient *L= Achievable Throughput / Demanded Throughput* ∈ [0, 1] is the ratio between the reinforcement and demand. However, for a high *L* the retrieved case had a successful action and $a_3$ is reused to address the current percept. In the case of low similarity, the percept is either stored or rejected, depending on KB memory design constraints. The agent behaves not purely by reflex but builds and exploits action-state representation from KB. This is a form of deterministic reasoning, while probabilistic reasoning may also be considered [15].

*Learning*: Learning is a property of the agent to improve its behavior based on experience, e.g. such that it can do more, it can do things better and/or it can do tasks faster, by improving action-state knowledge after applying an action. For wireless environment, active RL is considered to "intelligently guess" rich percepts to learn from and take new actions to probe these percepts given current information. In above example, the reinforcement such as user throughput indicates the quality of the taken action by learning coefficient *L* in KB or more efficiently by dynamic Q-learning previously introduced. Besides updating action-space values the selection of actions is needed. This is the task of an optimizer described in the next section. In some situations, when the user feedback is available semi-supervised learning such as SVM may be adopted to refine action-state values. Although learning is mainly about finding the best model that fits the data, it does not stand apart from the rest of AI disciplines.

For dynamic problems, strictly supervised learning techniques have issues with the lack of correct labels about the environment (i.e. dynamic traffic demand) leading to non-realistic models. Some of the other recent training-based methods are deep learning, meta-heuristic algorithms, fuzzy logic, genetic algorithms, HMMs, belief networks, etc. [4], [5].

*Optimizer*: The optimizer searches for new actions through exploitation and exploration functions in the second section. Given the agent functions in Fig. 6, the general idea behind optimization is to tune some of the agent components, e.g. KB, that are left unspecified to produce the required behavior. Optimizer is devising a search plan of action to achieve the goal by finding the best hypothesis within an action space [9]. To avoid keep selecting the "best" actions given what it knows the optimizer suggests actions that will lead to new and informative experiences given the reinforcement from interactions with environment as given in example above.

Exploration aims at discovering new search spaces that may lead to more promising solutions than the currently exploited solution set. Dynamic programming methods have been used in AI to solve optimization problems by storing intermittent actions so that they can be reused. Exploration policies adopted in wireless from other applications such as computer vision rely on Boltzmann or Gibbs distributions to randomly select an action. For example, to avoid service disruption due to random selection of frequency channels, the exploration needs to be controlled by problem-specific policy [12]. The selection within a set of given frequencies is well-known graph coloring NP-hard problem, where frequencies per nodes are selected to minimize interference, maximize load, etc. Other examples of actions are reconfiguration of adaptive modulation and coding, frame size, power and channel adjustment, antenna parameters, scheduling, handover parameters, routing and deployment location. More details about exploration/exploitation dilemma are presented in [12].

On the other side, exploitation greedily optimizes the network metrics within a limited search space in the KB (already experienced knowledge) that appears to be promising. For example, already used frequencies stored in the KB, corresponding to high action-state values, are reused when similar state of environment is perceived. This is simply done by adopting the action with maximum action-state value or Q-value. Thus, exploitation can be described as the one-step maximization of the expected reward.

---
[1] We only consider conceptual examples, while the detailed discussions (simulation and experimental validations) can be found in [11], [12].

*Decision-making*: Decision-making function evaluates the actions by the ability to meet the goal under the current and future percepts. Autonomous agent relies on on-line decision-making that comes up by training-free methods such as RL. Unlike supervised learning, where training data is available, in RL decisions are done in real time. For example, by using the learning coefficient in previous example the function checks both the similarity of states and the coefficient of the retrieved action-state. If an observation is not true, then the KB holds no matching case and two scenarios are possible: 1) a new case must be retained in the KB or 2) the best matching case has a suboptimal action that should be recomputed [11]. The optimization function is triggered to calculate a new action that will be executed and stored in the KB. We note that decision-making addresses the maximization of expected utility in episodic or sequential decision problems [15]. Some of the techniques for decision-making are Markov decision processes, game and optimization theory.

## Research Directions

We envision several research challenges as follows. Due to ultra-dense network deployments, the optimization function should consider multi-objective design strategy such as adversary learning, where reasoning function needs to consider other agents in a multiagent environment. The exploration of probabilistic reasoning and inference for network diagnostics based on belief or deep learning is an interesting challenge. More general problem of random exploration without negative impact on service delivery is interesting study. Beyond the fitting of data there are many issues such as domain knowledge representation; how when and what type of data to collect; and how to exploit the learned experiences to improve the agent functions. An efficient design of KB would be necessary in large-scale ultra-dense deployments where single or multiple instances of KB would be required. Transfer learning, where reusing of knowledge across different (physical) environments, is an open issue. Finally, integration of localization and user behavior data with AI framework may lead to improve the user experience.

## Conclusions

Unlike data-driven approaches, where knowledge-discovery supported by ML techniques is employed for prediction in matching problems, this paper presented a vision of autonomous wireless operations by knowledge management with AI disciplines such as sensing, reasoning and active learning. We presented basic principles and methodology for design of autonomous agent and discussed how to utilize AI disciplines in wireless systems.

| Self-organization feature | Characteristics | AI function | Models |
|---|---|---|---|
| Self-configuration | System configuration setup either on initial deployment of depending on the current critical situation in terms of network operations: cell coverage and deployment, neighbor cell list, authentication, maintenance updates, etc. Capability to maintain systems and devices depending on pre-defined system configuration. | Currently deterministic feature provided per network by operator's auto-configuration server or placed locally on memory as system configuration backup. Some ML models are applicable to automatically configure a set of parameters per cell to optimize local-policy. | K-means clustering, Hierarchical clustering, Fuzzy clustering |
| Self-optimizing | Deterministic (human rule-based) system checks with automated optimization of the local operation parameters according to global objectives: quality of service, capacity/bandwidth, coverage, etc. | Perception/Reasoning: deterministic - belief states are determined by logical formulas, e.g. classification | Dynamic programing, SVM, HMM, regression, RL, NN |
| | | Optimizer – constrained (convex) optimization functions not adaptable to network updates. | Optimization theory constraint minmax problems: linear programming, quadratic programming |
| | | Decision-making – rule based logical functions | Deterministic and not supported by learning |
| Self-healing | Machine-based system checks and methods for adapting configurations the system-of-systems: network and user's location-based updates. | Perception/Reasoning depending on a use case deterministic or probabilistic - belief is quantified as likely/unlikely or multi-class. | NN, HMM, SVM, Bayes networks |
| | | Optimizer – reinforcement learning, heuristics/meta-heuristics due to problem complexity supported by learning. | Policy-based Q-functions, multi-agents, game theory. |
| | | Decision-making supported by learning. | Optimization theory, game theory, MDPs, reinforcement learning |
| | | Learning – supervised, semi-supervised, unsupervised and reinforcement learning. | SVMs, ANNS, meta-heuristic algorithms, fuzzy logic, genetic algorithms, hidden Markov models, Belief networks, multi-agent learning |

*Table 1. Self-organization features and their relationship with different AI functions and models.*